\documentclass{aa}  

\usepackage{graphicx}
\usepackage{txfonts}
\usepackage{hyperref}
\hypersetup{
    colorlinks=true,
    linkcolor=blue,
    citecolor=blue,
    urlcolor=blue
    }
\usepackage{amsmath}	
\usepackage{siunitx}
\usepackage{pdflscape}
\usepackage{color}
\usepackage{xcolor}
\usepackage{ulem}

\usepackage{CJKutf8}
\newcommand{\CNnames}[1]{{\begin{CJK}{UTF8}{gbsn}~(#1)~\end{CJK}}}
\newcommand{\JPnames}[1]{{\begin{CJK}{UTF8}{min}~(#1)~\end{CJK}}}

\newcommand{\Teff}{{T_\mathrm{eff}}}

\newcommand{\logg}{{\log{(g)}}}
\newcommand{\loggf}{{\log gf}}
\newcommand{\logRpHK}{{\log{R'_\mathrm{HK}}}}
\newcommand{\He}{{He\,\num{10830}}}

\DeclareSIUnit\angstrom{\text {Å}}

\begin{document}

   \title{Stellar population astrophysics (SPA) with the TNG: Measurement of the \ion{He}{i} 10830\,\unit{\angstrom} line in the open cluster Stock 2
   \thanks{Based on observations made with the Italian Telescopio Nazionale Galileo (TNG) operated on the island of La Palma by the Fundación Galileo Galilei of the INAF (Istituto Nazionale di Astrofisica) at the Spanish Observatorio del Roque de los Muchachos. This study is part of the Large Program titled SPA – Stellar Population Astrophysics: the detailed, age-resolved chemistry of the Milky Way disc (A37TAC\_31, PI: L. Origlia), started in 2018 and granted observing time with HARPS-N and GIANO-B echelle spectrographs at the TNG.
   }
   }
   \titlerunning{\He~in Stock 2}
   \authorrunning{M. Jian et al.}


   \author{Mingjie Jian \CNnames{简明杰}\inst{1,2}
           \and
            Xiaoting Fu \CNnames{符晓婷}\inst{3,4}\and
            Noriyuki Matsunaga \CNnames{松永典之}\inst{2}\and
            Valentina D'Orazi\inst{5,6}\and
            Angela Bragaglia\inst{4}\and
            Daisuke Taniguchi\JPnames{谷口大輔} \inst{7}\and
            Min Fang \CNnames{房敏} \inst{3} \and
            Nicoletta Sanna\inst{8}\and
            Sara Lucatello\inst{6}\and
            Antonio Frasca\inst{9}\and
            Javier Alonso-Santiago\inst{9}\and
            Giovanni Catanzaro\inst{9}\and
            Ernesto Oliva\inst{8} 
          }

   \institute{
            Department of Astronomy, Stockholm University, AlbaNova University centre, Roslagstullsbacken 21, 114 21 Stockholm, Sweden\\
            \email{mingjie.jian@astro.su.se}
            \and
            Department of Astronomy, School of Science, The University of Tokyo, 7-3-1 Hongo, Bunkyo-ku, Tokyo 113-0033, Japan\and
            Purple Mountain Observatory, Chinese Academy of Sciences, Nanjing 210023, China \\\email{xiaoting.fu@pmo.ac.cn}
            \and
            INAF – Osservatorio di Astrofisica e Scienza dello Spazio di Bologna, via P. Gobetti 93/3, 40129 Bologna, Italy\and
            Department of Physics, University of Rome Tor Vergata, via della Ricerca Scientifica 1, 00133, Rome, Italy\and
            INAF - Osservatorio Astronomico di Padova, Vicolo dell' Osservatorio 5, 35122 Padova, Italy\and
            National Astronomical Observatory of Japan, 2-21-1 Osawa, Mitaka, Tokyo 181-8588, Japan\and
            INAF – Osservatorio Astrofisico di Arcetri, Largo E. Fermi 5, 50125 Firenze, Italy\and
            INAF–Osservatorio Astrofisico di Catania, via S. Sofia 78, 95123 Catania, Italy\\
             }

   \date{Received February 2, 2024; accepted April 13, 2024}

 
  \abstract
   {The precise measurement of stellar abundances plays a pivotal role in providing constraints on the chemical evolution of the Galaxy. 
    However, before spectral lines can be employed as reliable abundance indicators, particularly for challenging elements such as helium, they must undergo thorough scrutiny. 
    Galactic open clusters, representing well-defined single stellar populations, offer an ideal setting for unfolding the information stored in the helium spectral line feature. 
    In this study, we characterize the profile and strength of the helium transition at around \SI{10830}{\angstrom} (\He) in nine giant stars in the Galactic open cluster Stock 2. 
    To remove the influence of weak blending lines near the helium feature, we calibrated their oscillator strengths ($\loggf$) by employing corresponding abundances obtained from simultaneously observed optical spectra.
    Our observations reveal that \He~in all the targets is observed in absorption, with line strengths categorized into two groups. 
    Three stars exhibit strong absorption, including a discernible secondary component, while the remaining stars exhibit weaker absorption.
    The lines are in symmetry and align with or around their rest wavelengths, suggesting a stable upper chromosphere without a significant systematic mass motion.
    We found a correlation between \He~strength and \ion{Ca}{ii} $\logRpHK$ index, with a slope similar to that reported in previous studies on dwarf stars. 
    This correlation underscores the necessity of accounting for stellar chromosphere structure when employing \He~as a probe for stellar helium abundance. 
    The procedure of measuring the \He~we developed in this study is applicable not only to other Galactic open clusters but also to field stars,  with the aim of mapping helium abundance across various types of stars in the future.}
   \keywords{line:profiles -- stars: atmospheres -- stars: chromospheres -- stars: fundamental parameters -- (Galaxy:) open clusters and associations: individual: Stock 2
               }

   \maketitle
%

\section{Introduction}
\label{sec:intro}

Recent measurements of the stellar elemental abundance ratio for stars in our Milky Way has provided valuable insights into the formation and evolution of our Milky Way and its sub-systems.
Notably, studies have revealed that almost all globular clusters have multiple groups of stars with varying sodium and oxygen abundances, most probably because they formed in slightly separated epochs \citep[see e.g.][]{Gratton2019}, although some doubts on the reality of subsequent generations remain
 \citep[see e.g.][]{Bastian2018}.
There is also indirect evidence from photometry or CN line measurement that helium abundance varies within these stars \citep[see e.g.][]{Lee2005, Piotto2007, Bragaglia2010}.
Helium, as the second most abundant element in the Universe and the main product of hydrogen fusion in stellar nucleosynthesis, can serve as a tracer for the evolution of stars, clusters, and galaxies.

However, directly measuring He abundance of stars from their spectra is difficult. 
For stars hotter than $\sim$\SI{8500}{K}, a few helium lines form in their photosphere \citep[see e.g.][]{Villanova2009}. 
For cooler stars, determining the helium abundance is challenging due to the low temperatures (thus low thermal energy) in their photosphere, 
which prevent the formation of those lines since the populations of the lower levels of the corresponding transitions are small.

One particular helium spectral line, located at around \SI{10830}{\angstrom} (hereafter referred to as \He), 
corresponds to the transition between 2s and 2p triplet levels and can be observed in most late-type stars.
The triplet involves the excitation of an electron in the 2s$^3$S state of helium to one of the three excited states in the 2p$^3$P manifold, with the excitation potential $\sim$\SI{1.2}{eV} \citep{Ryabchikova2015}: \\
2s$^3$S -> 2p$^3$P$_0$ ($\lambda$ = \SI{10829.09}{\angstrom}),\\
2s$^3$S -> 2p$^3$P$_1$ ($\lambda$ = \SI{10830.25}{\angstrom}), \\
2s$^3$S -> 2p$^3$P$_2$ ($\lambda$ = \SI{10830.34}{\angstrom}).   \\
The strength of this \ion{He}{I} line, which is quantitatively represented by its equivalent width ($EW$), depends on the population of the lower state (2s).
As in cool stars, the photospheric radiation field has a too-low flux of highly energetic photons to efficiently populate the lower level of the transition, thus line must be formed in an outer atmospheric layer with a higher temperature, i.e. an upper chromosphere, and other excitation mechanisms must be at work. 
Previous studies (\citealt{Hirayama1971}, \citealt{Zirin1975}, \citealt{Zarro1986} and \citealt{Avrett1994}) suggested the photoionization recombination (PR) mechanism, in which helium atoms are ionized by high-energy photons and subsequently recombine to the triplet states, can populate the lower state of \He.
On the other hand, collisional excitation (CE) can directly excite atoms from singlet levels to triplet levels if the electron temperature in the atmosphere ($T_\mathrm{e}$) exceeds \SI{20000}{K} \citep{Athay1956, Andretta1997}.
In the case of a CE regime, the strength of \He~is expected to be related to the chromosphere structure, which can be probed by diagnostics of the chromosphere as the cores of \ion{Ca}{ii} H\&K lines.
Moreover, we expect a correlation with diagnostics of magnetic activity at diverse atmospheric layers, such as the X-ray luminosity, which traces the coronal emission.

The complex formation mechanisms involved make it challenging to establish the relationship between the strength of \He~and stellar parameters, particularly the helium abundance (e.g., \citealt{Dupree2013}). 
Consequently, it remains difficult to accurately measure the helium abundance using the strength of the \He~transition. Therefore, there is an urgent need to investigate and determine the elusive connection between the \He~strength and other stellar parameters to enable spectroscopic measurements of helium abundance in cool stars.


Many studies have been based on observations of \He~of cool stars and investigated its strength.
Some early investigations (e.g., \citealt{Vaughan1968}, \citealt{Zirin1976}, \citealt{Smith1983} and \citealt{Obrien1986})  acquired low-resolution spectra of \He~for a substantial number of stars.  \citet{Zirin1982} established an empirical relation between the line strength and various stellar parameters, including effective temperature ($\Teff$), \ion{Ca}{ii} K line intensity, and X-ray luminosity.
These findings were subsequently confirmed by studies such as \citet{Zarro1986} and \citet{Takeda2011}.
With the development of near-infrared (NIR) high-resolution spectrographs, \He~studies have entered the high-resolution era. The resolution around the \He~line can reach $\Delta \lambda$\,$\sim $\,\SI{0.2}{\angstrom} or even smaller, allowing for much more detailed investigations of spectral features
\citep[see e.g.][]{Smith2004, Sanz-Forcada2008, Dupree2009, Pasquini2011, Smith2012}. 
However, variations in measurement techniques, such as profile fitting versus direct pixel summation, and the presence or absence of telluric line corrections, may introduce systematic or statistical uncertainties when comparing results from different studies.   
Therefore, it is crucial to reexamine the empirical relations established over forty years ago by \citet{Zirin1982}  with high-resolution spectrographs.

For the relation between \He~line strength and \ion{Ca}{ii} H\&K line emissions ($\lambda \sim$ \num{3931}--\SI{3970}{\angstrom}),  obtaining spectra that simultaneously cover these lines is crucial to account for possible variations in the level of stellar activities and avoid potential biases.
In this study, marking the first step of our probing helium abundance work series, we establish a method to accurately measure the strength and profile of the \He~line based on high-resolution spectra obtained from the GIARPS mode \citep{Claudi2017} of the Telescopio Nazionale Galileo (TNG) is ideally suited. We then apply this method to a set of similar stars within a single stellar population, specifically red clump stars in the open cluster Stock 2.

Stock 2 is a nearby open cluster discovered by \citet{Stock1956}, at a distance of \SI{400}{pc} \citep{Dib2018}.
Its age and chemical composition were not well known until recently.
The first detailed spectroscopic analysis for this cluster was reported by \citet{Reddy2019}.
With a sample of three giants, they estimated the cluster's iron abundance ([Fe/H]) value to be $-0.06 \pm 0.03$.
\citet{Ye2021} obtained a similar value of $-0.04 \pm 0.15$ from LAMOST medium-resolution spectra of around 300 stars.
As a part of the TNG Large Programme Stellar Population Astrophysics \citep[SPA, see e.g., ][]{Origlia2019, Casali2020, Zhang2021}, \citet{Alonso-Santiago2021} analysed the spectra of 32 dwarfs, 14 main-sequence turnoff stars and ten giants, and derived their stellar parameters and chemical abundances.
They conclude that the age and iron abundance of the cluster are \SI{450}{Myr} and [Fe/H]=$-0.07 \pm 0.06$.
As the member stars share similar ages and chemical compositions, this is an ideal sample for studying the behaviour of helium lines in a given stellar population.

We outline the observations and explain the process of data reduction in Sect.~\ref{sec:data}.
Section~\ref{sec:blending} focuses on characterizing the blending of \He. 
In Sect.~\ref{sec:line-measurelent}, we describe the methods used to measure \He~and \ion{Ca}{ii} H\&K lines. 
Finally, we present the result and discussion in Sect.~\ref{sec:result}.

\section{Data and reduction}
\label{sec:data}

The data used in this study consist of high-resolution NIR and optical spectra of 9 giants in the open cluster Stock 2, obtained as a part of the SPA project.
The data used here for Stock 2 were observed in the GIARPS mode at the TNG. 
GIARPS combines the GIANO-B \citep{Oliva2006} for obtaining NIR spectra (\num{0.9}--\SI{2.45}{\micro m}) with a resolution of $R=\num{50000}$, and the HARPS-N \citep{Cosentino2012} for obtaining optical spectra (\num{0.383}--\SI{0.690}{\micro m}) with $R=\num{115000}$, simultaneously capturing the required spectral ranges.
The simultaneous observation of \ion{Ca}{ii} H\&K lines using HARPS-N and \He~lines with GIANO-B provides a direct comparison of their behaviour. This is particularly valuable as it enables the avoidance of any potential time-dependent variations between these lines.

\begin{table*}
    \caption{Observation log: the notation, name, observation date in Reduced Heliocentric Julian date ($\mathrm{RHJD} = \mathrm{HJD} - 2400000$), Gaia DR2 $G$ band absolute magnitude $M_G$, $G_{BP} - G_{RP}$ colour, exposure time and the signal-to-noise ratio per pixel ($\mathrm{S/N_\He}$) at the continuum level around \He~of our sample stars.}
    \centering
    \begin{tabular}{ccccccc}
    \hline
    Star & Name & RHJD & $M_G$ & $G_{BP} - G_{RP}$ &$t_\mathrm{exp}$\,(s) & $\mathrm{S/N_\He}$ \\
    \hline\hline
      g1 & HD\,15498 & 58428.407 & -0.884 & 1.487 & 700 & 73\\
      g2 & HD\,14346 & 58428.390 & -0.565 & 1.463 & 700 & 103\\
      g3 & HD\,13437 & 58428.468 & -0.371 & 1.443 & 1400 & 89\\
      g4 & HD\,13207 & 58428.450 & 0.214 & 1.473 & 1400 & 78\\
      g5 & HD\,14403 & 58428.487 & 0.430 & 1.262 & 1400 & 79\\
      g6 & HD\,12650 & 58428.423 & 0.310 & 1.591 & 1400 & 75\\
      g7 & HD\,15665 & 58429.341 & 0.392 & 1.449 & 1400 & 73\\
      g9 & HD\,13655 & 58429.359 & 0.744 & 1.711 & 1400 & 60\\
     g10 & HD\,13134 & 58429.378 & 0.928 & 1.468 & 1400 & 74\\
    \hline
    \end{tabular}
    \label{tab:stock-2-obslog}
\end{table*}

\begin{figure}
	\centering
	\includegraphics[width=1\columnwidth]{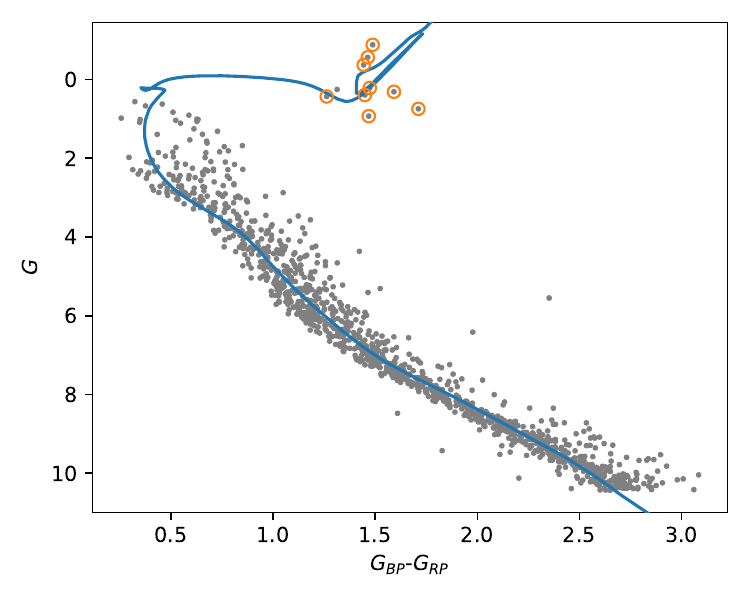}
	\caption{Colour-magnitude diagram in Gaia DR2 photometry bands of Stock 2 member stars (grey) from \citet{Cantat-Gaudin2018}, with the sample stars in this study marked in orange circles. The PARSEC isochrone with an age of \SI{450}{Myr}, [M/H] = -0.07 and extinction of $A_\mathrm{V} = 0.84$ is overplotted in blue.}
	\label{fig:stock2-CMD}
\end{figure}

The age, iron abundance (as a proxy of metallicity), differential reddening, and chemical abundances of Stock 2 stars are reported in \citet{Alonso-Santiago2021} based on the HARPS-N data of our GIARPS observations.
Here we follow the naming convention from Table~1 of \citet{Alonso-Santiago2021} and present the target list in Table~\ref{tab:stock-2-obslog}, as well as the stellar parameters in Table~\ref{tab:stock-2-para}.
Fig.~\ref{fig:stock2-CMD} shows the Gaia DR2 colour-magnitude diagram of our targets, along with the other member stars from \citet{Cantat-Gaudin2018}.
The PARSEC isochrone version~1.2S\,\footnote{obtained from the CMD 3.7 input form \url{http://stev.oapd.inaf.it/cgi-bin/cmd}} (\citealt{Bressan2012}) with an age of \SI{450}{Myr}, a metallicity of [M/H]=\num{-0.07}, and an extinction of $A_V = 0.84$ is overplotted as a reference.
We select the Gaia DR2 photometric system from the PARSEC CMD 3.7 webpage for the isochrone  with the bolometric correction from YBC \citep{Chen2019}.
The isochrone parameters are adopted from \citet{Alonso-Santiago2021}.
Note that the scatter of the red clump (RC) stars and the extended main sequence are mainly due to the differential reddening in this cluster \citep[see the detailed discussions in][]{Alonso-Santiago2021}. 

While we lack precise asteroseismic data to robustly distinguish between red giant branch (RGB) and RC stars, we are confident that our sample stars are most likely RC stars. This confidence stems not only from their placement within a relatively narrow range of colour and absolute magnitude, but also from the probabilistic nature of observations.
The turn-off mass of Stock 2 is about 2.8\,M$_{\odot}$ \citep{Alonso-Santiago2021}. Stars with this mass evolve very fast along the sub-giant branch and red giant branch, making them challenging to capture in observation samples.
In contrast,  stars in this mass range spend a more considerable amount of time in the RC stage. As a result, these RC stars have a substantially higher probability of being sampled in observations.  
This increased sampling probability, combined with the constraints of colour and absolute magnitude, strongly suggests that our sample stars are predominantly RC stars.

\begin{figure}
    \centering
    \includegraphics[width=1\columnwidth]{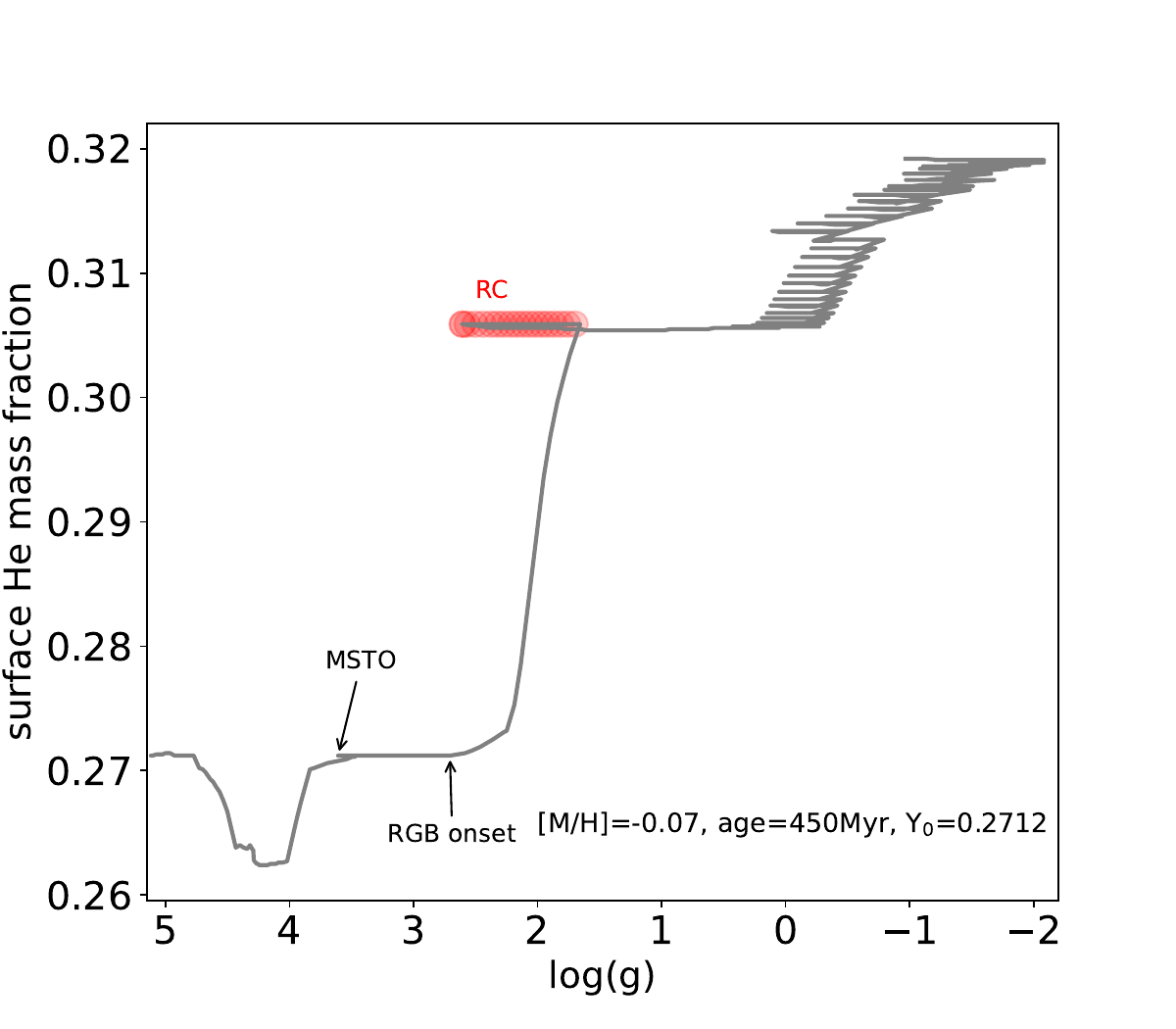}
    \caption{He mass fraction (Y) on the stellar surface as a function of surface gravity $\logg$ for the Stock 2 isochrone. The initial He mass fraction of this isochrone is Y$_0$=0.2712. The locations of the MSTO, RGB onset,  and the RC phase are highlighted.
    }
    \label{fig:he_logg}
\end{figure}

We aim to measure \He~in RC stars because their surface helium abundance are expected to be the same.
This expectation arises from the assumption that within a single stellar population, such as an open cluster, Stock 2 in our case, the surface helium abundance remains relatively consistent for RC stars.
In Fig.~\ref{fig:he_logg} we present the surface He mass fraction (Y) in a single stellar population characterized by the metallicity of Stock 2.
The PARSEC isochrone used here aligns with the one employed in Fig.~\ref{fig:stock2-CMD}, which includes the surface helium content evolution.
To gauge the stage of stellar evolution, we use the surface gravity $\logg$ as an indicator.
During the main sequence phase, the surface helium mass fraction Y decreases (see evolutionary phase with $\logg$ $\gtrapprox$  4 in the figure) due to microscopic diffusion.
The Microscopic diffusion leads the surface abundance decrease for elements heavier than hydrogen.
For the microscopic diffusion treatments in stellar models and its impact on the surface elemental abundances, we recommend readers to the PARSEC modelling paper \citep[e.g.][]{Bressan2012, fu2015} for more detailed discussions.
This decrease in Y is later counterbalanced by the first dredge-up at the end of the main sequence, restoring Y to its initial value, $Y_0$. This restoration is evident in the range of $\logg \approx 4$ to the main sequence turn-off (MSTO).
The surface He abundance of main sequence stars in Stock 2 is strongly affected by the microscopic diffusion. However, giant stars in Stock 2, with an approximate mass of  2.8\,M$_{\odot}$, undergo a negligible microscopic diffusion during their main sequence, attributed to their narrow surface convective zone. 
In the subsequent sub-giant branch phase, from the MSTO to the RGB onset, the surface He content remains consistent with the initial value $Y_0$.
The onset of the RGB phase is marked by the merging of the hydrogen-burning shell and the stellar surface convective zone, leading to a rapid increase in surface Y due to the transportation of fresh helium from the hydrogen-burning shell. 
This increase continues up to the tip of the RGB.
In the subsequent RC phase (highlighted by red-filled dots in the figure), stars share a similar stellar structure and exhibit a nearly constant surface helium content. 
This uniformity in surface He abundance means that variations in the \He~ line observed in these RC stars  predominantly reflect the local chromospheric conditions, where this spectral line is formed. 
While sub-giant phase stars also exhibit similar surface helium levels, their rapid evolution makes them challenging to sample in observations. 
As main sequence and RGB stars in the cluster experience either a decrease or increase in surface He abundance, measuring \He~ in RC stars effectively eliminates additional uncertainties that might arise from variations in the He abundance itself.

\begin{figure*}
	\centering
	\includegraphics[width=2\columnwidth]{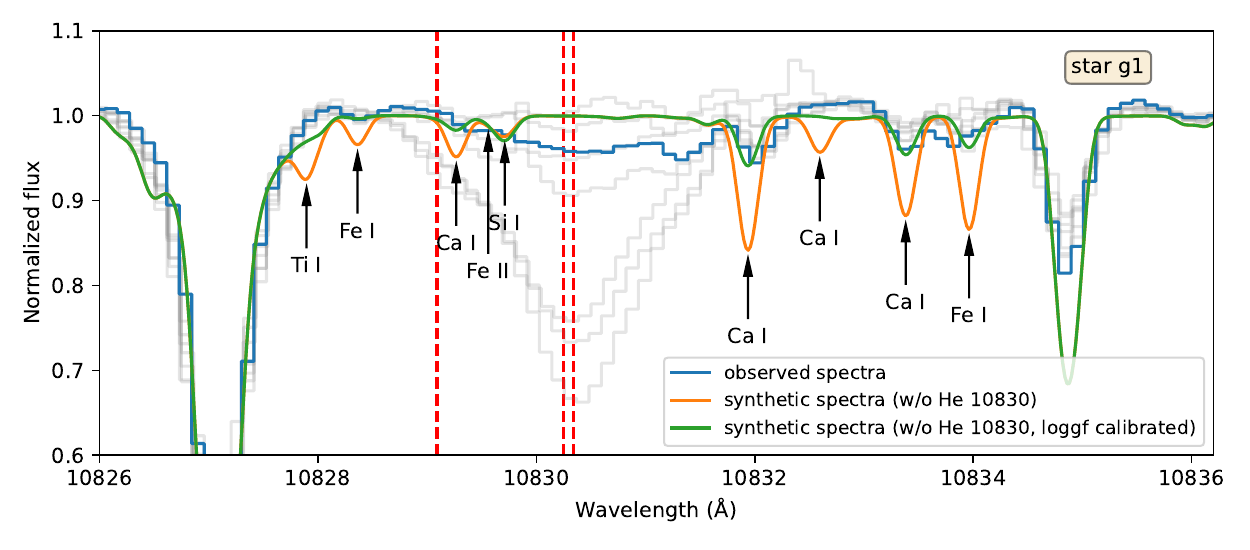}
	\caption{The observed (blue step line; with telluric correction) and the synthetic spectra before (orange) and after (green) their $\loggf$ values calibrated (without the \He) for star g1. Red vertical dash lines indicate the position of the \He~triplet, and the black arrows indicate the position of blending lines with $\loggf$ values calibrated (see Sect.~\ref{sec:blending}). The spectra of the other eight target stars are plotted in grey as a reference.}
	\label{fig:stock2-spec-example}
\end{figure*}

The reduction of the GIANO-B spectra was carried out using the data reduction pipeline software GOFIO \citep{Rainer2018}.
This pipeline applied the needed corrections, i.e., the removal of bad pixels and cosmic rays, subtraction of sky and dark frame, and correction of flat-field and blaze, to the observed frames.
The normalized 1D spectra were then extracted by the pipeline and wavelength calibrated.
The telluric lines, i.e., the spectral lines that originated from the earth's atmosphere instead of the stellar atmosphere, are also present in these extracted spectra.
We used a python package, \texttt{telfit}\footnote{\url{https://telfit.readthedocs.io/en/latest/}} \citep{Gullikson2014}, to fit the telluric lines and then subtract them from the observed spectra.
A new continuum normalization was performed after the subtraction.
These corrected spectra were used for the measurement of the stellar and line parameters in Sect.~\ref{sec:blending}.
We note that such a procedure of telluric correction may alter the shape of \He~in some cases. 
Thus for the measurement of \He, instead of using the telluric corrected spectra, all lines around \SI{10830}{\angstrom} were fitted, including the telluric lines (see Sect.~\ref{sec:he-measurement} for more detail).
Fig.~\ref{fig:stock2-spec-example} shows an example of \He.
To evaluate the stellar activity of our sample stars, we use the same set of HARPS-N spectra as in the research by \citet{Alonso-Santiago2021}.
The normalization process and line measurement will be described in Sect.~\ref{sec:logRpHK-measurement}.

\begin{table*}
    \centering
    \caption{Effective temperature ($\Teff$), surface gravity ($\logg$) and [Fe/H] measured from the HARPS-N spectra in \citet{Alonso-Santiago2021}, as well as the microturbulence velocity ($\xi$), [Fe/H] and broadening velocity ($V_\mathrm{broad}$) in this study using GIANO-B spectra.}
    \begin{tabular}{ccccccc}
    \hline
    Star & $T_\mathrm{eff}, \mathrm{K}$ & $\log{g}$ & [Fe/H] & $\xi, \mathrm{km\,s^{-1}}$  & [Fe/H] & $V_\mathrm{broad}, \mathrm{km\,s^{-1}}$ \\
     & (HARPS-N) & (HARPS-N) & (HARPS-N) & (GIANO-B) & (GIANO-B) & (GIANO-B)\\
    \hline\hline
      g1 &  $4530\pm86$ & $2.14\pm0.10$ &  $0.01\pm0.09$ & $1.57\pm0.20$ & $-0.12\pm0.05$ & $4.6\pm1.1$ \\
      g2 & $4760\pm111$ & $2.69\pm0.14$ &  $0.02\pm0.10$ & $1.10\pm0.24$ & $-0.04\pm0.05$ & $6.3\pm1.6$\\
      g3 & $4937\pm114$ & $2.51\pm0.35$ &  $0.04\pm0.08$ & $1.32\pm0.26$ & $-0.01\pm0.06$ & $6.3\pm1.5$\\
      g4 & $4977\pm117$ & $2.82\pm0.18$ &  $0.04\pm0.08$ & $1.21\pm0.19$ &  $0.03\pm0.04$ & $4.7\pm1.4$\\
      g5 &  $5061\pm56$ & $2.99\pm0.19$ &  $0.04\pm0.07$ & $1.07\pm0.24$ &  $0.02\pm0.05$ & $5.8\pm1.2$\\
      g6 & $5002\pm110$ & $2.96\pm0.20$ &  $0.03\pm0.07$ & $0.87\pm0.20$ &  $0.09\pm0.04$ & $4.7\pm1.2$\\
      g7 &  $5058\pm56$ & $2.97\pm0.20$ &  $0.03\pm0.07$ & $1.39\pm0.31$ &  $0.02\pm0.06$ & $6.3\pm1.0$\\
      g9 &  $5062\pm56$ & $3.00\pm0.19$ &  $0.00\pm0.09$ & $1.18\pm0.24$ & $-0.01\pm0.05$ & $5.6\pm1.2$\\
     g10 &  $5066\pm56$ & $3.01\pm0.19$ & $-0.03\pm0.09$ & $0.97\pm0.29$ & $-0.02\pm0.05$ & $5.6\pm1.3$\\
    \hline
    \end{tabular}
    \label{tab:stock-2-para}
\end{table*}

\section{Blending around \He}
\label{sec:blending}

The presence of line blending, where spectral lines overlap in wavelength, can significantly alter the shape and strength of our target line, particularly when it is wide and weak.
If we measure the strength of \He~without accounting for the blending from the nearby spectral lines, the line strength would be over-estimated. 
Consequently, determining the extent of line blending around the \He~for our target stars is necessary before accurately measuring the strength of \He~can be achieved.
By comparing observed spectra with synthetic spectra, we can disentangle the contributions of \He~and other spectral lines, allowing for a more precise evaluation of the blending effects.
Fig.~\ref{fig:stock2-spec-example} presents the synthetic spectra around \He~(excluding the helium feature) for star g1 in our sample. 
The synthetic spectra are generated using the LTE radiative transfer code MOOG (version November 2019; \citealt{Sneden2012}) with the ATLAS9 stellar atmosphere model from \citet{Castelli2003} and the line list from Vienna Atomic Line Database (VALD3 without hyperfine structure; \citealt{Ryabchikova2015})\footnote{They are integrated into the python wrapper \texttt{pymoog} \citep{pymoog}.}.
The VALD3 database provides a comprehensive line list on our target wavelengths.

The prominent atomic lines around \He~are labelled with black arrows in  Fig.~\ref{fig:stock2-spec-example}.
The figure clearly illustrates the presence of numerous atomic lines, such as \ion{Ca}{i}, \ion{Fe}{i}, \ion{Si}{i}, which contaminate the \He~line region. Therefore, it is crucial to account for these lines when measuring the $EW$ of \He~to ensure accurate results. Their contributions need to be carefully considered and taken into account in the analysis.

The observed spectra reveal that these lines are weaker than they appear in the synthetic spectra. 
This discrepancy exceeds the range of errors associated with the parameters such as $\Teff$, $\logg$, metallicity, or elemental abundances. 
The most likely cause of this difference is the incorrect atomic parameters of the line, particularly the $\loggf$ in the infrared region.
As demonstrated by \citet{Andreasen2016} with the solar spectrum, incorrect $\loggf$ values obtained from VALD3 can result in more than one dex discrepancies in the abundances of certain NIR Fe lines, even when using the same solar atmosphere model parameters. 
To address this issue, they proposed a list of calibrated $\loggf$ values based on solar abundances. 
The blending lines around the \He~region in our sample stars face a similar challenge. 
These lines' $\loggf$ values need to be calibrated using the observed spectra before accurately assessing their contribution to the \He~line. 
The high resolution and high signal-to-noise ratio of the GIANO-B spectra obtained for Stock 2 makes them an ideal sample for such calibration efforts.

In this section, we will start our calibration by determining the global stellar parameters which affect the line shape and strength, mainly microturbulence velocity of a Gaussian profile, $\xi$, and broadening velocity, $V_\mathrm{broad}$, then using the elemental abundance derived from the HARPS-N spectra, which are observed simultaneously, to calibrate the $\loggf$ values of each blending line.


\subsection{Measurement of $\xi$ and broadening velocity}

Microturbulence velocity  $\xi$ represents the small-scale non-thermal motion in the stellar atmosphere.
It mainly affects the strength of the saturated lines.
Broadening velocity is an indicator of the combination of larger-scale non-thermal motion, or macroturbulence, and stellar rotation.
These two parameters, along with $\Teff$, $\logg$, metallicity and elemental abundances, are the main factors that alter the spectral line.
\citet{Alonso-Santiago2021} determined the values of $\Teff$, $\logg$, [Fe/H] and elemental abundances for our sample stars using the ROTFIT code.
However, ROTFIT does not yield values for microturbulence velocity ($\xi$). Instead, it adopts the microturbulence and macroturbulence from template spectra, and calculates the projected rotational velocity ($v\sin i$).
To conduct a detailed characterisation of the \He~ line features, in this work we determine $\xi$ and the broadening velocity $V_\mathrm{broad}$ for our sample stars using the GIANO-B spectra. 

We follow the approach described in \citet{Kondo2019} to determine $\xi$ and $V_\mathrm{broad}$.
It is summarized in two steps:

\begin{enumerate}
    \item Select isolated \ion{Fe}{i} lines in the GIANO-B wavelength range.
    \item Alter $\xi$, [Fe/H] and $V_\mathrm{broad}$ to fit the observed spectra using MPFIT algorithm \citep{Takeda1995}.
\end{enumerate}

In the first step, the isolated \ion{Fe}{i} lines are defined as those with clear detectable absorption (depth $d > 0.05$) at the signal-to-noise ratio ($S/N_\mathrm{\He}$) of our sample and a small overlap with any other eventual line, whose depth must be less than 20\% at the line centre. 
We avoid the regions with strong telluric absorption, such as \num{11060}--\SI{11550}{\angstrom}, \num{13150}--\SI{15150}{\angstrom}, \num{17600}-\SI{21000}{\angstrom} and $>\SI{23500}{\angstrom}$.
Such criteria yield 179 \ion{Fe}{i} lines from \num{9820} to \SI{22473}{\angstrom}.

In the second step, the observed spectra (those with the instrumental broadening from GIANO) for all the lines selected in step one are fit by choosing the $\xi$ value so that the line-by-line [Fe/H] is not dependent on the line strength. 
The fitting of $V_\mathrm{broad}$ is added during this process to provide a better fit between the observed and synthetic spectra.
[Fe/H] is also determined in step two, which provides a good comparison to the value determined from the optical (HARPS-N) spectra.

\begin{figure*}
	\centering
	\includegraphics[width=2\columnwidth]{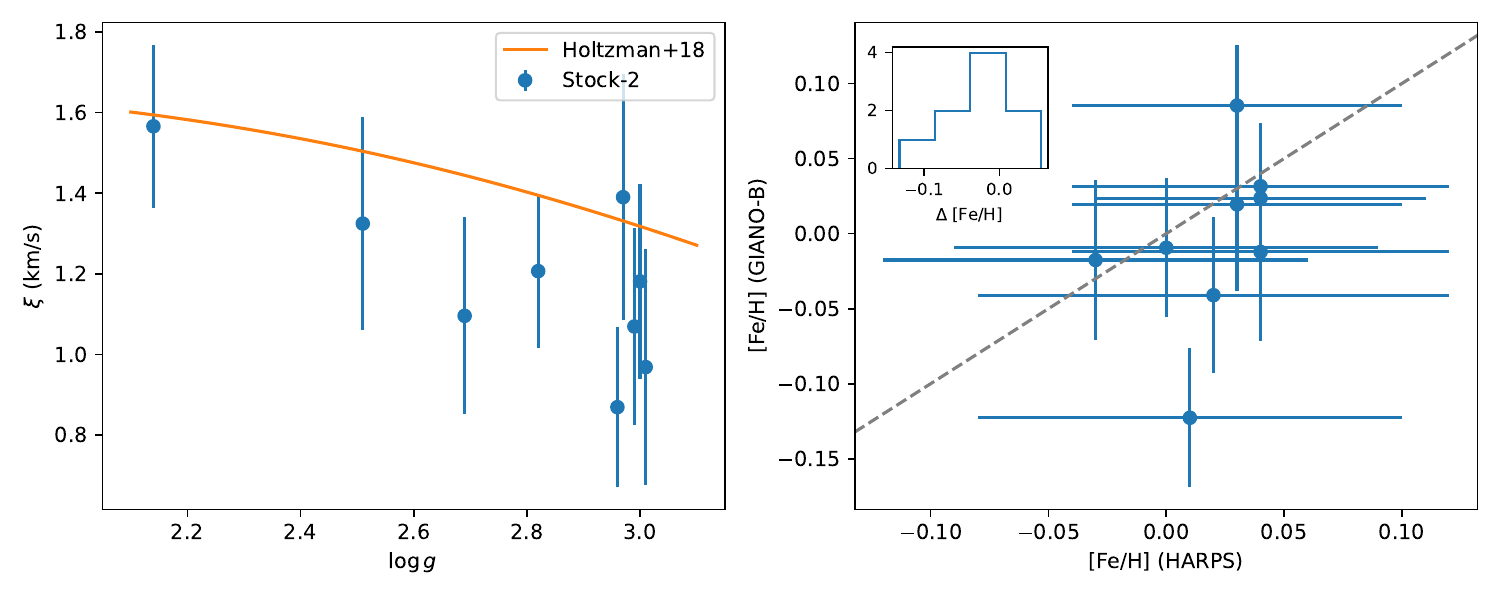}
	\caption{Left panel: The measured $\xi$ from MPFIT versus $\logg$ for the Stock 2 stars, along with the relation reported in \citet{Holtzman2018}. Right panel: the [Fe/H] determined from GIANO-B spectra versus those from HARPS-N \citep{Alonso-Santiago2021}, with the histogram of $\Delta\mathrm{[Fe/H]}\equiv \mathrm{[Fe/H]_{\operatorname{GIANO-B}}-[Fe/H]_{\operatorname{HARPS-N}}}$ in the upper-left corner.}
	\label{fig:stock2-vmi_mh}
\end{figure*}

A bootstrap method was then used to determine the parameters together with their uncertainties for each star.
We chose 179 lines randomly from those selected in step (i) but allowed repeat selection (i.e., some lines are selected more than one time while others may be discarded), and apply MPFIT using these lines for a star.
This procedure is then repeated 1000 times, and the final values, as well as the errors of $\xi$, [Fe/H], and $V_\mathrm{broad}$ are set as the median and standard deviation of the obtained results.
Table~\ref{tab:stock-2-para} lists these parameters for our sample.
The derived $\xi$ values are consistent with the trend reported by \citet{Holtzman2018} from the APOGEE giants, and the precision of the metallicities measured using the GIANO-B spectra is similar to those obtained from the HARPS spectra, as shown in Fig.~\ref{fig:stock2-vmi_mh}.
The measured $V_\mathrm{broad}$ are around \SI{5.5}{km\,s^{-1}}.
This is in perfect agreement with the low values found by \citet{Alonso-Santiago2021}, who measured a $v \sin{i}$ less than \SI{4}{km\,s^{-1}} (on average) for the giants in Stock 2.

\subsection{$\loggf$ calibration of blending lines}
\label{sec:loggf-calib}

With all the global parameters that affect the line strength determined in the above section, the $\loggf$ values of the blending lines around \He~are now ready to be calibrated. 
The strongest lines in the synthetic spectrum near \He, which are indicated by black arrows in Fig.~\ref{fig:stock2-spec-example} and reported in Table 3, were selected to be calibrated.
Other lines are found to have a negligible contribution using the synthetic spectra, with a maximum depth of less than 0.01 for our sample stars.
The strong \ion{Si}{i} line at \SI{10827.088}{\angstrom} on the left side of \He~was not included for calibration due to its potential impact on the wing of \He.
It will be treated during the fitting of \He~(Sect.~\ref{sec:he-measurement}).

The $\loggf$ values of these lines were then adjusted to find the best fit between the synthetic and observed spectra, starting with the strongest line.
Here we adopted the Si, Ca, Ti and Fe abundance ratios from \citet{Alonso-Santiago2021}.
Calibration results were discarded if the line was affected by remaining telluric absorption, i.e., the \ion{Fe}{i} line at \SI{10833.96}{\angstrom}, or if they are affected by \He, i.e., the \ion{Ca}{i}, \ion{Fe}{ii} and \ion{Si}{i} lines at around \SI{10829.50}{\angstrom}.
The final $\loggf$ value for each line is set as the median of the $\loggf$ values from all the stars.
The values before and after the calibration are presented in Table~\ref{tab:loggf_blending} and the synthetic spectra for star g1 are shown in Fig.~\ref{fig:stock2-spec-example}.
Most of the calibrated values are smaller than the original ones in the VALD line list.



\begin{sidewaystable*}
    \caption{centre wavelength, element and ionization state, excitation potential (EP), van der Waals damping parameter (C6), and oscillation strength ($\loggf$)  from VALD3 database, together with the calibrated $\loggf$  for each star in our sample, and the final adopted $\loggf$ values. The NaN value  ``...'' indicates those with the calibration result discarded (see Sect.~\ref{sec:loggf-calib}).}
    \centering
    \begin{tabular}{ccccc|cccccccccc}
    \hline
    wavelength &   Ele.  &    EP &   C6 &  $\loggf$ & \multicolumn{10}{c}{calibrated~ $\loggf$}     \\
    (\unit{\angstrom}) & Ion & (\unit{eV}) & & VALD3 & g1 & g2 & g3 & g4 & g5 & g6 & g7 & g9 & g10 & final   \\
    \hline\hline
    10831.94 &  \ion{Ca}{i} & 4.878  & -6.940 & -0.029&       -0.329 &       -0.398 &       -3.629 &       -1.141 &       -0.529 &       -3.629 &       -0.579 &       -0.529 &        -1.104 &      -0.579 \\
    10833.97 &  \ion{Fe}{i} & 5.587  & -7.540 & -1.223&           ... &       -2.023 &           ... &       -2.123 &           ... &       -1.836 &       -1.523 &       -1.898 &            ... &      -1.898 \\
    10833.38 &  \ion{Ca}{i} & 4.877  & -7.590 & -0.244&       -0.488 &       -1.094 &       -1.119 &       -0.738 &       -1.044 &       -1.494 &       -0.619 &       -0.794 &        -0.619 &      -0.794 \\
    10829.27 &  \ion{Ca}{i} & 4.441  & -7.090 & -1.224&       -1.574 &           ... &       -1.524 &       -2.324 &           ... &       -1.724 &       -1.824 &           ... &            ... &      -1.724 \\
    10828.36 &  \ion{Fe}{i} & 5.446  & -7.520 & -2.104&       -2.654 &       -4.804 &       -2.754 &       -2.454 &       -5.004 &       -2.504 &       -3.104 &       -5.204 &        -3.104 &      -3.104 \\
    10829.71 &  \ion{Si}{i} & 6.727  & -6.860 & -1.765&       -1.565 &           ... &           ... &       -4.465 &           ... &           ... &       -1.665 &           ... &            ... &      -1.665 \\
    10832.60 &  \ion{Ca}{i} & 4.878  & -6.930 & -0.742&       -4.042 &       -3.642 &       -4.042 &       -3.842 &       -3.842 &       -3.642 &       -4.042 &       -3.842 &        -3.642 &      -3.842 \\
    10829.56 & \ion{Fe}{ii} & 5.585  & -7.871 & -3.312&       -3.212 &           ... &           ... &       -3.512 &           ... &           ... &       -5.712 &           ... &            ... &      -3.512 \\
    10827.90 &  \ion{Ti}{i} & 0.836  & -7.810 & -3.910&       -6.810 &       -6.610 &       -6.610 &       -6.410 &       -5.810 &       -6.710 &       -6.010 &       -6.210 &        -6.010 &      -6.410 \\
    \hline
    \end{tabular}
    \label{tab:loggf_blending}
\end{sidewaystable*}

\subsection{Blending $EW$ across the Kiel diagram}

\begin{figure}
    \centering
    \includegraphics[width=1\columnwidth]{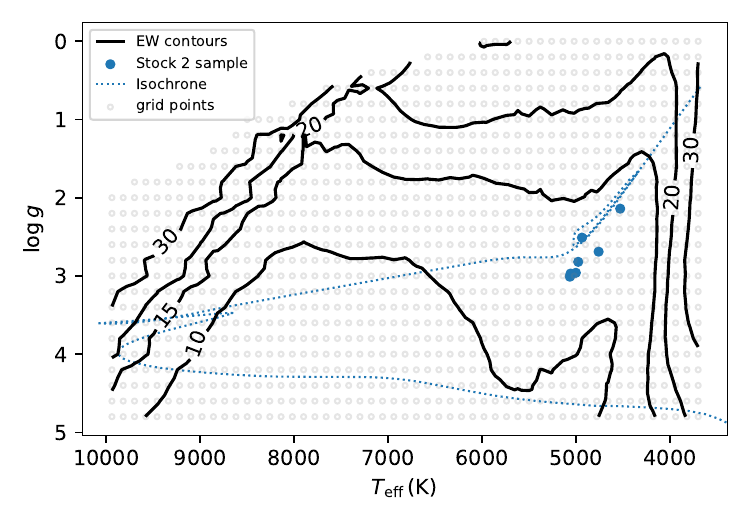}
    \caption{
    The $EW$ contours (in \unit{m\angstrom}) of blending lines across the Kiel diagram, i.e. $\Teff$--$\logg$, in solar metallicity and GIANO-B resolution ($R=50000$). The grey circles are the grid points used to calculate the contours, and the blue points indicate our Stock 2 sample. The isochrone used in Fig.~\ref{fig:stock2-CMD} is plotted as a reference.
    }
    \label{fig:stock2-blending-EW}
\end{figure}

We use the calibrated $\loggf$ values of the blending lines to estimate the blending around \He. 
The $EW$ of blending lines between \num{10828.60} and \SI{10831.50}{\angstrom} was calculated from synthetic spectra at the GIANO-B resolution across the Kiel  diagram ($\Teff$ vs. $\logg$), using the calibrated line list.
The contour of blending $EW$s is shown in Fig.~\ref{fig:stock2-blending-EW}. 
For the Stock 2 sample stars, the blending $EW$s are around \SI{13}{m\angstrom}.
The blending across the Kiel diagram is systematically smaller for dwarfs, i.e., higher $\logg$, and larger for supergiants.
We note that the contour may change at another resolution or metallicity, i.e., blending will become larger for lower resolution or higher metallicity, but overall it is small compared to the \He.
Thus the measured $EW$ of \He~will be subtracted by the blending $EW$ in the following analysis using a wavelength range where the depth of fitted \He~feature is larger than 0.005.

\section{Measurement of the line parameters}
\label{sec:line-measurelent}

In this section, we will outline the method for measuring the parameters of \He~and \ion{Ca}{ii} H\&K lines. 
The parameters of \He~include the $EW$ and shape of the feature, while the measurement of the \ion{Ca}{ii} H\&K lines primarily focuses on the core emission, represented by the $\logRpHK$ index.

\subsection{Equivalent width and profile measurement for \He}
\label{sec:he-measurement}

\begin{figure*}
	\centering
	\includegraphics[width=2\columnwidth]{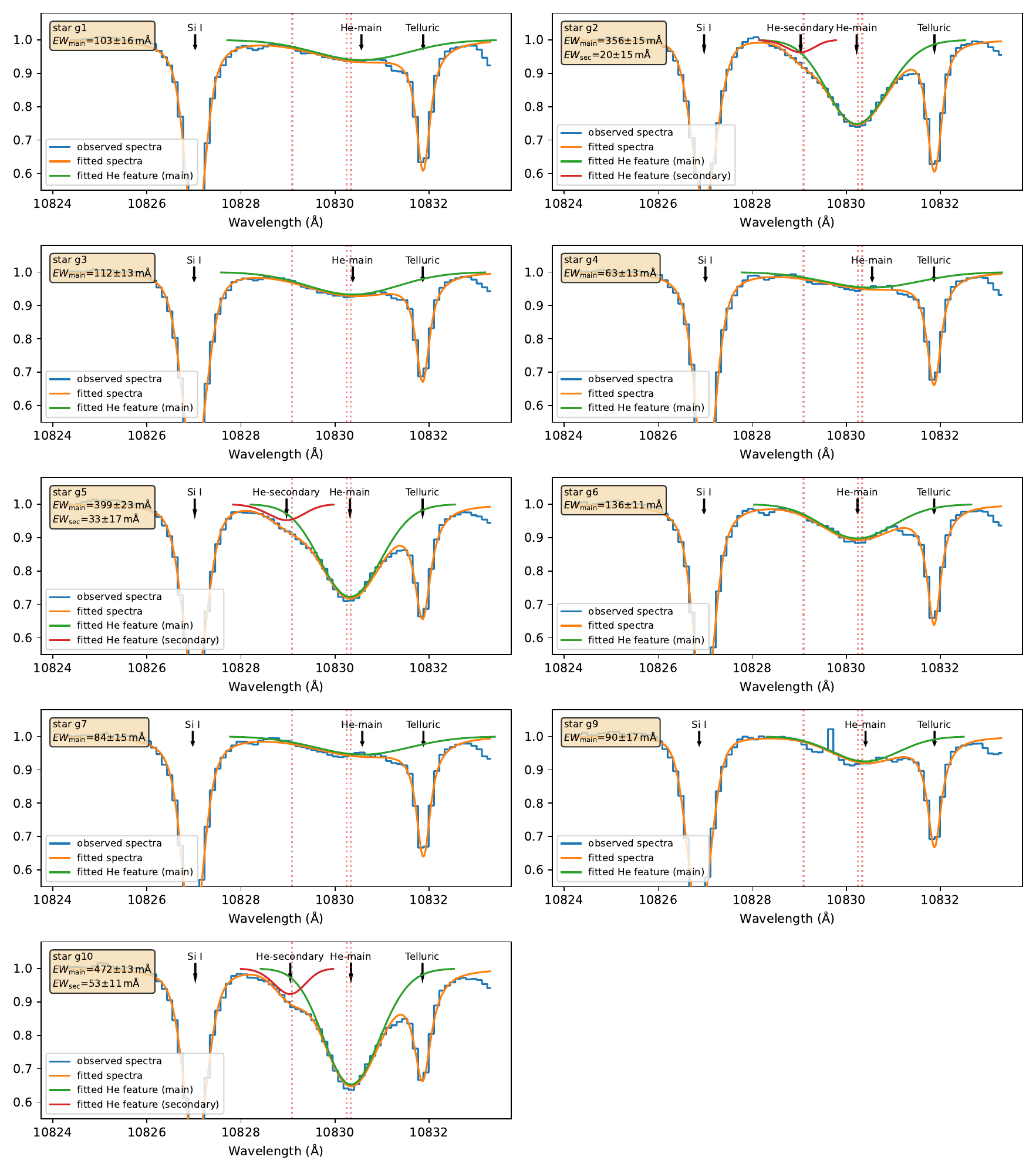}
	\caption{The observed (blue; without telluric correction) and fitted spectra (orange) for our sample stars. The main and secondary components of \He~are fitted as the green and red curve, respectively.}
	\label{fig:EW-all}
\end{figure*}

As \He~is a broad absorption feature, it is more susceptible to the data reduction procedures such as telluric correction and continuum normalization. 
The closest telluric line near \He~is the one at \SI{10832.10}{\angstrom}, which overlaps with the right wing of \He~in most of our spectra (Fig.~\ref{fig:EW-all}).
A strong \ion{Si}{i} line is also present on the left side.
To maintain the integrity of \He's profile during our measurement, we use the spectra with telluric lines and include them in the fitting procedure, as stated at the end of Sect.~\ref{sec:data}.

Fig.~\ref{fig:EW-all} presents the fitting of the features around \He.
A skew Gaussian profile is used to fit \He. 
It is defined as:
\begin{equation}
    sG(x; A, \mu, \sigma, \gamma) = \frac{A}{\sigma\sqrt{2\pi}} \left\{ 1+\mathrm{erf}\left[\frac{\gamma(x-\mu)}{\sigma\sqrt{2}}\right]\right\} \exp{\left[-\frac{(x-\mu)^2}{2\sigma^2}\right]}
\end{equation}
where $A$, $\mu$ and $\sigma$ are the amplitude, centre and width of the distribution, $\gamma$ represent its skewness, and $\mathrm{erf}$ is the error function:
\begin{equation}
    \mathrm{erf}(x) = \frac{2}{\sqrt{\pi}} \int_0^x e^{-x^2} \mathrm{d}x.
\end{equation}

Such a profile allows for the asymmetry to be accounted for, if any, by the extra $\gamma$ parameter in addition to a Gaussian profile.
To reproduce other prominent components, i.e., the \ion{Si}{i} line at \SI{10827.09}{\angstrom} and the telluric line at \SI{10834.00}, two Voigt profiles $V(x; x_0, a, \gamma, \sigma)$\footnote{defined as $V(x; x_0, a, \gamma, \sigma) = \int_{-\infty}^\infty \frac{1}{\sigma \sqrt{2\pi}} \mathrm{e}^{-\frac{x'^2}{2\sigma^2}} \frac{a\gamma}{\pi[(x-x_0-x')^2] + \gamma^2} \mathrm{d}x'$} are used for fitting, with the centre wavelength, Lorentzian amplitude, Lorentzian and Gaussian full width at half maximum represented by $x_0$, $a$, $\gamma$ and $\sigma$ respectively.
The continuum is fitted using a first-order polynomial. 
In the spectra of stars with relatively strong \He~absorption, i.e., g2, g5 and g10, the weaker component of \He~triplet, located at around \SI{10829}{\angstrom} is also visible. 
Another skew Gaussian profile is added to the fitting of these spectra, and the stronger and weaker components are named as main and secondary in the following analysis.

The parameters for \He, including the $EW$, full-width at half-maximum (FWHM) and line centre wavelength ($\lambda_\mathrm{centre}$) are derived from the fitted skew Gaussian profile(s). 
To quantify the asymmetry of the profile, we calculate the ratio of the area on the left and right side of $\lambda_\mathrm{centre}$ in \He~feature as $B/R$, the blue-to-red ratio. 
The error of these parameters was estimated from the Monte-Carlo method. 
We generated 1000 mock observed spectra using the best-fit parameters and $\mathrm{S/N_\He}$, then measured them in the same way we did for the observed spectra.
The errors of the fitted parameters were then derived from the standard deviation of those measured from the mock spectra. 


\begin{sidewaystable*}
    \centering
    \caption{Measurement results of the sample stars. The $EW$, FWHM, $\lambda_\mathrm{centre}$ and $B/R$ for the main and secondary component of \He~and $\logRpHK$ are listed.}
    \begin{tabular}{c|cccc|cccc|c}
    \hline
    Star & $EW_\mathrm{main}$ & $\mathrm{FWHM_{main}}$ & $\lambda_\mathrm{centre, main}$ & $(B/R)_\mathrm{main}$ & $EW_\mathrm{sec}$ & $\mathrm{FWHM_{sec}}$ & $\lambda_\mathrm{centre, sec}$ & $(B/R)_\mathrm{sec}$ & $\log{R'_\mathrm{HK}}$ \\
     & (m\AA) & (\AA) & (\AA) & & (m\AA) & (\AA) & (\AA) & & \\
    \hline
    g1 & $103\pm16$ & $2.36\pm0.19$ & $10830.56\pm0.12$ &  $0.99\pm0.02$ &       ... &           ... &               ... &                ... &   $-5.42$ \\
    g2 & $356\pm15$ & $1.52\pm0.05$ & $10830.22\pm0.11$ &  $0.92\pm0.04$ & $20\pm15$ & $0.71\pm0.26$ & $10829.03\pm0.11$ &      $1.08\pm0.05$ &   $-4.62$ \\
    g3 & $112\pm13$ & $2.28\pm0.15$ & $10830.38\pm0.09$ &  $0.99\pm0.02$ &       ... &           ... &               ... &                ... &   $-5.15$ \\
    g4 &  $63\pm13$ & $2.37\pm0.23$ & $10830.55\pm0.15$ &  $1.00\pm0.02$ &       ... &           ... &               ... &                ... &   $-5.14$ \\
    g5 & $399\pm23$ & $1.53\pm0.07$ & $10830.32\pm0.14$ &  $0.96\pm0.04$ & $33\pm17$ & $0.91\pm0.28$ & $10828.97\pm0.14$ &      $1.07\pm0.04$ &   $-4.85$ \\
    g6 & $136\pm11$ & $1.79\pm0.12$ & $10830.24\pm0.05$ &  $0.94\pm0.03$ &       ... &           ... &               ... &                ... &   $-5.05$ \\
    g7 &  $84\pm15$ & $2.35\pm0.20$ & $10830.58\pm0.12$ &  $0.98\pm0.02$ &       ... &           ... &               ... &                ... &   $-5.44$ \\
    g9 &  $90\pm17$ & $1.69\pm0.24$ & $10830.41\pm0.09$ &  $0.99\pm0.04$ &       ... &           ... &               ... &                ... &   $-5.17$ \\
    g10 & $472\pm13$ & $1.40\pm0.04$ & $10830.34\pm0.06$ &  $0.92\pm0.04$ & $53\pm11$ & $0.79\pm0.13$ & $10829.05\pm0.06$ &      $1.07\pm0.04$ &   $-4.58$ \\
    \hline
    \end{tabular}
    \label{tab:stock-2-res}
\end{sidewaystable*}

\subsection{Measurement of \ion{Ca}{ii} lines}
\label{sec:logRpHK-measurement}

The core emission of the \ion{Ca}{ii} H and K lines is commonly quantified through the $R'_\mathrm{HK}$ index:
\begin{equation}
    R'_\mathrm{HK} = \frac{F'_\mathrm{H} + F'_\mathrm{K}}{\sigma T^4_\mathrm{eff}},
\end{equation}
where $F'_\mathrm{H}$ and $F'_\mathrm{K}$ represent the integrated absolute flux (in the unit of \SI{}{erg / s / cm^2}) at the stellar surface for the bands located in the centres of the H and K line after excluding those emitted in the photosphere.
The S-index, defined as the ratio of the normalized flux between the triangle windows at the centre of \ion{Ca}{ii} H\&K lines and two windows on both sides of the lines \citep{Wilson1968}, can also be used to quantify the \ion{Ca}{ii} H\&K core emission.
However, it contains a photospheric contribution at the line core and is also found to be sensitive to the colour of the star (see, e.g., Fig. 1 and 2 in \citealt{Noyes1984}).
$R'_\mathrm{HK}$ then can better represent the condition in the stellar chromosphere compared to S-index, and we will limit ourselves to $R'_\mathrm{HK}$ in the following analysis.

\begin{figure*}
	\centering
	\includegraphics[width=2\columnwidth]{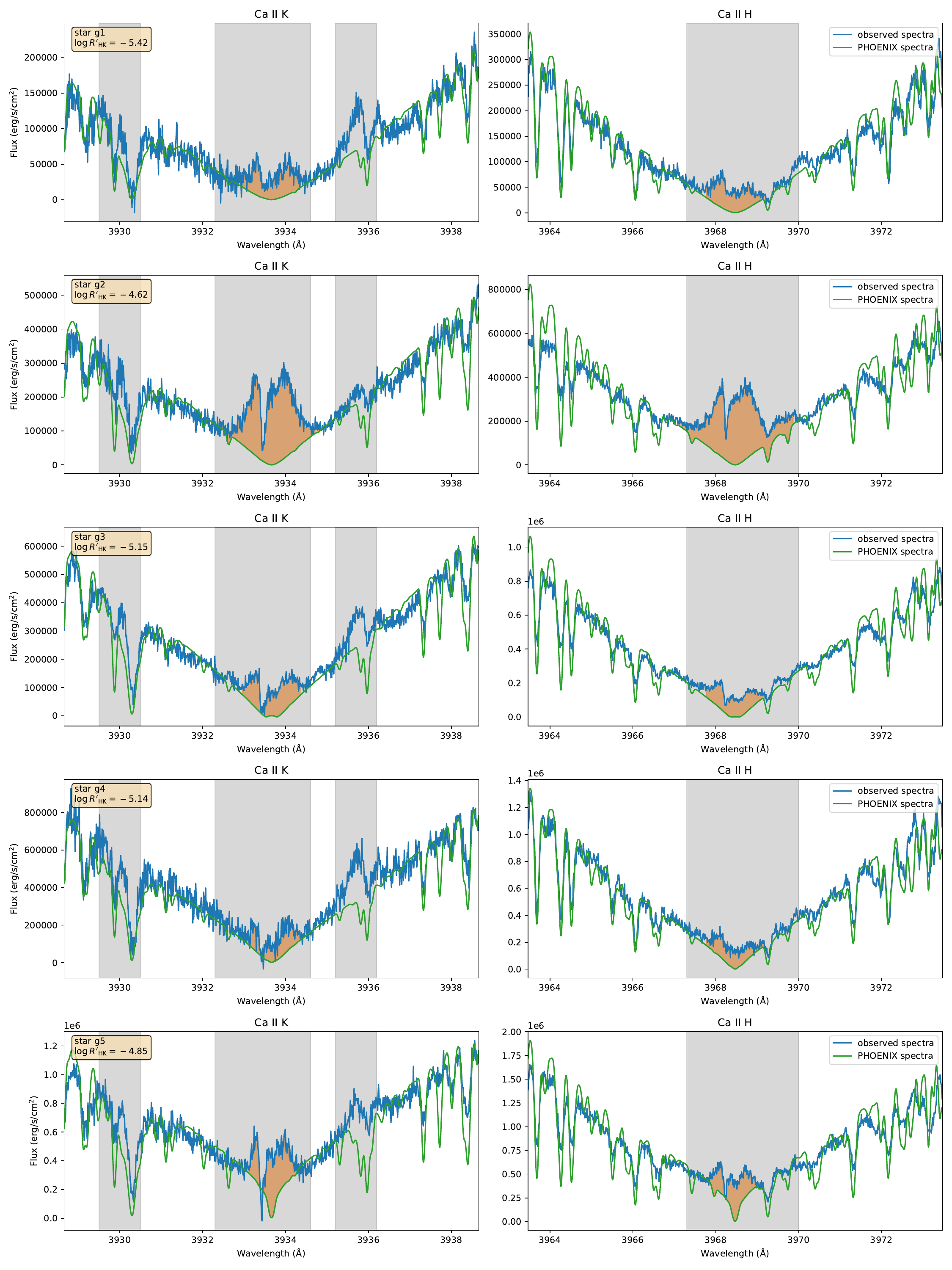}
    \caption{The \ion{Ca}{ii} H\&K spectra for stars g1--5, scaled to the PHOENIX scale. The vertical grey strips indicate the lines masked out before the spline fitting, and the orange shaded areas in the middle of the lines are used for measuring the $\logRpHK$ index.}
	\label{fig:CaII-example-1}
\end{figure*}

\begin{figure*}
	\centering
	\includegraphics[width=2\columnwidth]{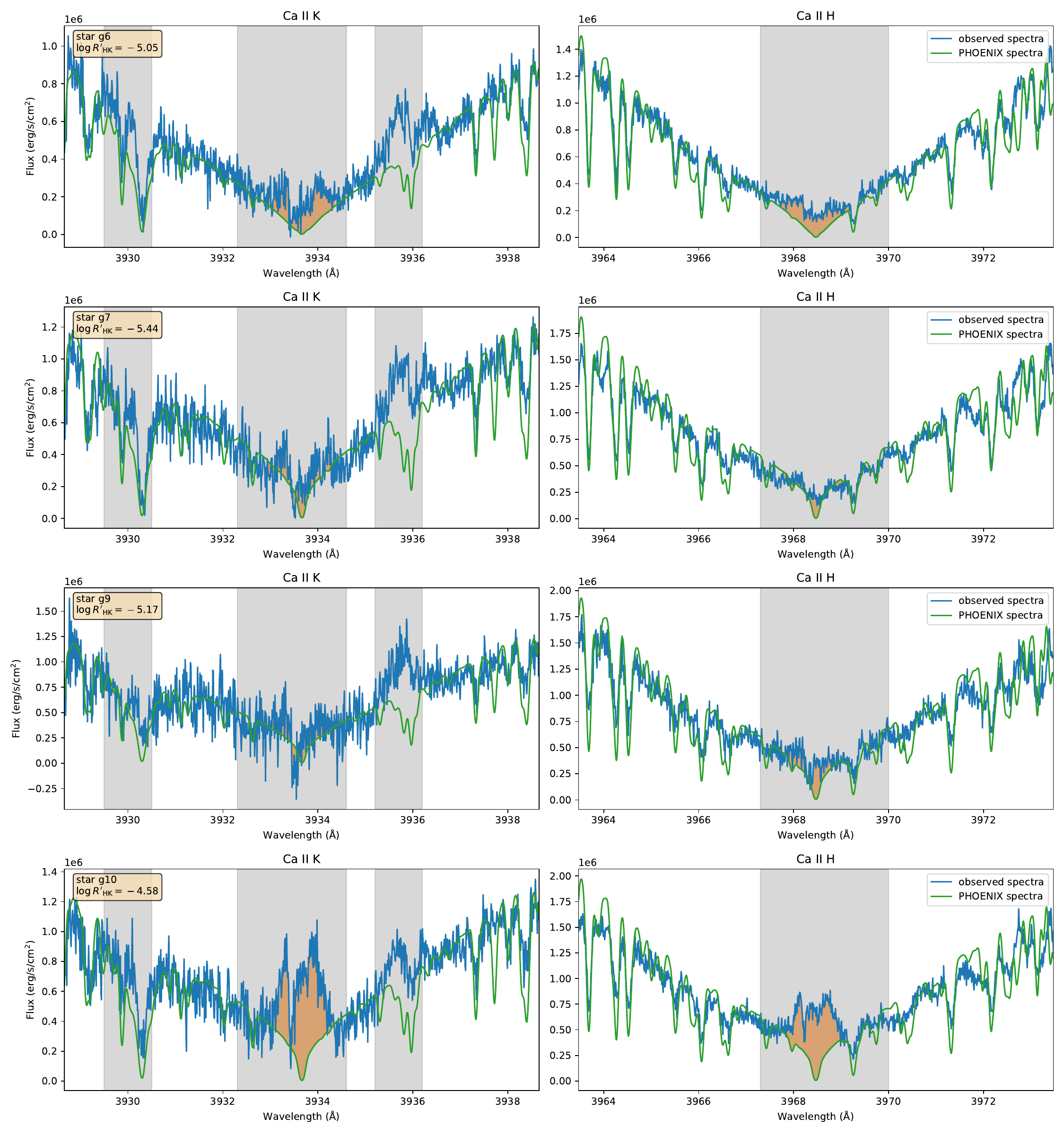}
    \caption{Same as Fig.~\ref{fig:CaII-example-1}, but for star g6, g7, g9 and g10.}
	\label{fig:CaII-example-2}
\end{figure*}

Our method of measuring the $R'_\mathrm{HK}$ index is described as follows.
The photospheric flux at the centre of \ion{Ca}{ii} H\&K are adopted from the synthetic spectra.
As shown in Fig.~\ref{fig:CaII-example-1} and \ref{fig:CaII-example-2}, we first extract high-resolution synthetic spectra with no chromospheric contribution from the PHOENIX database \citep{Husser2013} using Starfish\footnote{\url{https://github.com/Starfish-develop/Starfish}}, which are then broadened and resampled to match the spectral resolution of our sample.
The observed spectra are then vertically scaled by fitting a spline function to a 60-pixel running median to the ratio between the observed and model flux in the wavelength range between \num{3921} and \SI{3989}{\angstrom}.
The strong and broad spectral lines and the core of \ion{Ca}{ii} H\&K lines are masked out before the spline fitting.
Finally, $F'_\mathrm{K}$ and $F'_\mathrm{H}$ are measured from the difference between the observed and synthetic spectra in the line cores of the K and H bands wherever there is an obvious difference between the observed and synthetic spectrum (the orange areas at the line centres in Fig.~\ref{fig:CaII-example-1} and \ref{fig:CaII-example-2}), from which the $R'_\mathrm{HK}$ indices are then determined.
This procedure is similar to the one adopted by \citet{Frasca2023} for measuring \ion{Ca}{ii}\,H\&K fluxes for the members of ASCC123 observed with HARPS-N.

The final \He~measurement results, including the $EW$, FWHM, line centre wavelength, and the blue-to-red ratio of the primary and secondary components, along with the $\logRpHK$ index measurement results, are summarised in Table~\ref{tab:stock-2-res}.
The $EW$s of the main \He~feature exhibit a range from \num{84} to \SI{471}{m\angstrom}, and their corresponding errors are approximately \SI{10}{m\angstrom}, thanks to the high spectral resolution and high S/N.
All of the \He~features in our sample are in the form of absorption and can be easily seen in all the spectra (see Fig.~\ref{fig:EW-all}).

\section{Result and Discussion}
\label{sec:result}

\subsection{Two He components}

The nine giants in our sample are separated into two groups: Stars g2, g5 and g10 have a relatively large $EW$, i.e, greater than \SI{350}{m\angstrom}, while others have $EW$ around \SI{100}{m\angstrom}.
The three stars with stronger \He~absorption also show a secondary absorption component around the wavelength $\lambda$ = \SI{10 829.088}{\angstrom} (see their spectra in Fig.~\ref{fig:EW-all}). 
The $EW$ of the secondary \He~component is around 5--10\% of their main component.

The existence of the secondary component has been reported in the literature, though mainly for main sequence stars \citep[see e.g.][]{Takeda2011, Andretta2017}.
In \citet{Takeda2011} the second \He~component is displayed (see their Fig. 2) but not measured because the \SI{10 829.088}{\angstrom} line is considerably weaker than the other two He lines. 
This component is called ``the minor component'' in \citet{Andretta2017} for the X-ray dwarf stars. Its $EW$, if can be fitted independently, is around \num{15}--\num{25}\% of the main component, which is larger than the ratios we measured in our giant sample stars.
\citet{Andretta2017} suggest that the ratio should be 1:8 (12.5\%) in optically thin condition according to Russell–Saunders coupling, and different ratios refer to the different optical thicknesses of the He line components.
Future observations of stars covering different evolutionary stages but similar stellar populations may shed light on the discrepancy between our results on giant stars and their results on dwarfs.
For instance, understanding whether the possible difference in the chromosphere structure between dwarf and giant is the main factor affecting the strength ratio.

\subsection{\He~line asymmetry}

Given that the \He~lines originate from the upper chromosphere of the stars, their line asymmetry provides insights into the local environment's motion.
In our sample stars, the \He~line shapes do not exhibit any evidence of bulk motion in their upper chromosphere. 
The line centre wavelengths of both the main and the second components are corrected for the RVs of each star (adopted from the Tab.~2 of \citealt{Alonso-Santiago2021}). 
Following this correction, the centre wavelengths of the main components are closely aligned with their rest wavelengths.
A minor red-shift is detected, corresponding to a velocity of \num{0}--\SI{9}{km/s}.
This subtle shift indicates the absence of significant systematic mass motions in the region where the \He~ line forms, i.e., upper chromosphere.
According to the results presented in Table~\ref{tab:stock-2-res}, the $B/R$ ratio is found to be very close to 1 for both the main and second components of \He. This result suggests that both components exhibit a high degree of symmetry.

An asymmetrical \He~absorption line is not always the case for giant stars.
\citet{Dupree1992} studied the \He~profile of a few field stars including  HD6833, $\alpha$ Boo, and $\alpha$ Aqr, and found that their \He~present a P Cygni profile, i.e. emission He line with absorption on the blueward side.
They concluded that such a profile of HD6833 indicates a very strong mass loss from the star, with a motion velocity of at least \SI{90}{km\,s^{-1}}, and the other stars show similar profiles.

Conversely, the symmetric \He~profile exhibited by our sample stars, closely aligned with or around their rest wavelengths, suggests that the region in which the line forms -- the upper chromosphere -- experiences negligible or small bulk motion.
This is not surprising because our sample stars are very likely in the red clump stage, which is less affected by mass loss \citep{Girardi1999}.
The stars in  \citet{Dupree2009} are more active than our sample, since they are located higher on the red giant branch and stronger outflows are expected.
We note that, however, a larger sample of stars with \He~measurement is necessary to validate the slightly red-shifted trend in our sample stars.

\subsection{The $EW(\mathrm{He})$-$\logRpHK$ relation}

\begin{figure}
	\includegraphics[width=1\columnwidth]{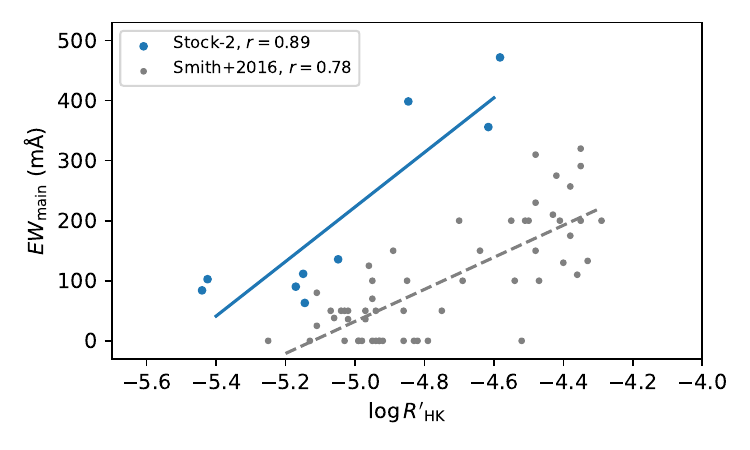}
	\caption{$EW(\mathrm{He})$ of the main \He~component versus $\logRpHK$ for Stock 2 RC stars. For comparison, dwarf stars from \citet{Smith2016} are plotted in grey. The $r$ values indicate the Pearson correlation coefficients of the linear fitting for the RC and dwarf stars.}
	\label{fig:stock-2-Ca-He}
\end{figure}

We observe a correlation between the strength of the main \He~component and the $\logRpHK$ index.
As depicted in Fig.~\ref{fig:stock-2-Ca-He}, stars exhibiting a less pronounced core emission in the \ion{Ca}{ii} H\&K lines, denoted by $\logRpHK \lesssim -5$, also display weaker helium absorption in their spectra. 
Conversely, the three stars characterized by larger  $EW$ and a secondary He component demonstrate stronger \ion{Ca}{ii} H\&K core emission ($\logRpHK > -5$). 
This observation suggests a connection between stronger stellar activity and the enhanced strength of the \He~feature.

A similar positive relation between the strength of \He~and $\logRpHK$ was reported in \citet{Smith2016} for the Galactic field dwarf and sub-dwarf stars. Their sample stars are predominantly population I, with a small number of population II dwarfs.
For comparison, we over-plot their results in Fig.~\ref{fig:stock-2-Ca-He} with grey dots and fit a linear line to show their relation with $EW$ and the $\logRpHK$ index. 
Stars in \citet{Smith2016} with $\logRpHK \sim -5$ show a weak absorption of $EW \sim \SI{50}{m\angstrom}$, whilst those with $\logRpHK \sim -4.5$ show $EW \sim \SI{200}{m\angstrom}$.

Dwarf and sub-dwarf stars, as observed in \citet{Smith2016}, consistently exhibit smaller \He~$EW$ systematically compared to RC stars of Stock 2 studied in this work. 
This disparity primarily stems from differences in surface helium abundance between dwarf and RC stars, under the assumption of similar initial metallicity and helium abundance. 
We delved into this aspect in our sample selection process, as detailed in Section \ref{sec:data}, and illustrated in Fig.~\ref{fig:he_logg}.
Additionally, it is worth noting that variations in resolution and differences in measurement methods, such as how the continuum was determined, between the two studies, could also contribute to the observed disparity in $EW$.

The variation of the \He~$EW$ in the Stock 2 RC stars, together with the strong positive correlation between the \He~$EW$ and $\logRpHK$ shown in  Fig.~\ref{fig:stock-2-Ca-He}, suggests that the chromosphere structure affects the \He~feature, and it needs to be checked before using \He~to probe the stellar helium abundance.
A possible approach to check the chromosphere structure is observing \ion{Ca}{ii} H\&K lines and \He~simultaneously, then either comparing the \He~with stars showing similar \ion{Ca}{ii} H\&K core emission \citep{Pasquini2011}, or determining the chromosphere structure from \ion{Ca}{ii} H\&K lines and fitting \He~by varying the helium abundance \citep{Dupree2013}.

\subsection{Comparison with previous \He~measurements}

\begin{figure}
	\includegraphics[width=1\columnwidth]{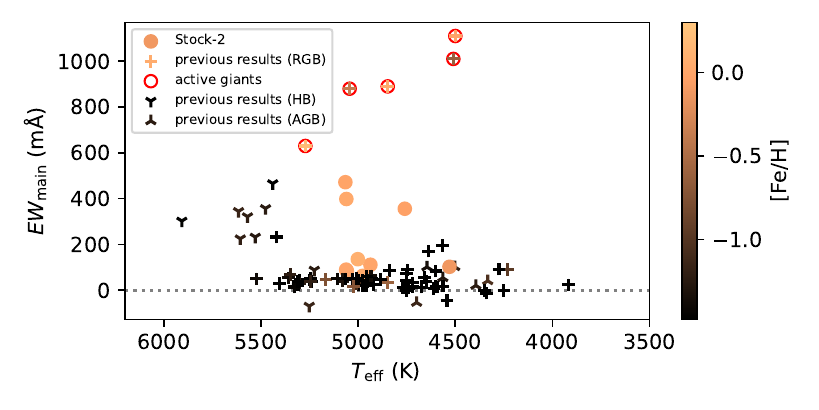}
	\caption{$EW\mathrm{(He)}$ versus $\Teff$ for the Stock 2 stars with giants from previous studies. The $EW$s smaller than 0 indicate that the \He~is in emission.}
	\label{fig:he_compare_stock-2}
\end{figure}

Our $EW$ measurement of the Stock 2 sample is compared with those obtained in previous studies, as illustrated in Fig.~\ref{fig:he_compare_stock-2}.
The data from literature encompasses field red giants sourced from \citet{Sanz-Forcada2008}, \citet{Dupree2009} and \citet{Takeda2011}, in addition to giants hailing from the globular cluster $\omega$ Cen (\citealt{Dupree2011} and \citealt{Navarrete2015}).
It is worth noting that these stars either exhibit significant activity levels (in the case of stars from \citealt{Sanz-Forcada2008}) or are characterized by their metal-poor nature as globular cluster members.
The \He~strengths of the active giants in the literature are $\sim$$\SI{800}{m\angstrom}$, while the other giants with lower [Fe/H] (dark colour in the figure) have a much weaker absorption, around \SI{50}{m\angstrom}.
The \He~strength of the asymptotic giant branch stars is also $\sim$$\SI{50}{m\angstrom}$, while that of the horizontal branch stars are stronger for higher $\Teff$, as reported in \citet{Strader2015}.

The strength of \He~in our sample lies in between the very active giants and metal-poor giants from previous studies.
The $EW$ of stars with weaker \He~absorption is slightly larger than the metal-poor giants, while the three giants with the largest $EW$ have weaker features than the active giants.
Since most of the giants from previous studies have a [Fe/H] lower than \num{-1}, the slightly larger $EW$ of our sample may imply that the metal-rich stars have a slightly stronger \He~than inactive metal-poor stars.
Due to the difference in how the \He~is measured in all these studies, a homogeneous measurement of \He~using the same method with a larger sample with [Fe/H] from \num{-1} to solar for both dwarfs and giants would yield a better conclusion on the possible metallicity effect.

\subsection{Future applications}

The method we developed in this study to measure the strength and profile of \He~can be applied to other datasets.
The high resolution of GIANO-B spectrograph enables the $\loggf$ calibration for the blending lines, which quantifies the blending line $EW$, not only for the current dataset, but also for other stars across the HR diagram.
The difference between the calibrated and original $\loggf$ values suggests that the line parameters in this wavelength are not accurate and need to be calibrated, or that other effects are playing a role in the formation of these lines. 
\He~is well sampled in the observed spectra, which provides information on the profile of this feature. 
In a following paper we will  apply this method to other open clusters observed in the SPA large programme.
The method  can also be applied to spectra at lower resolutions, such as those observed with the WIDE mode of WINERED, the Warm INfrared Echelle spectrograph to Realize Extreme Dispersion and sensitivity ($R\approx 28000$; \citealt{Ikeda2022}).
The data from spectrographs which can cover \ion{Ca}{ii} H\&K and \He~simultaneously, e.g., X-SHOOTER \citep{Vernet2011}, will also provide vital information on the behaviour of \He.
Such observation is especially useful for globular cluster giants, whose helium abundances are expected to be varied.
In short, we foresee the application of our method to the data obtained by various spectrographs across various types of stars in future works. 

\section{Summary}
\label{sec:summary}


In this paper, we present a pilot study of the helium content in the Stock 2 red clump stars using the \He~spectral line.
Understanding stellar helium abundance is crucial for comprehending both stellar evolution and the chemical evolution of the Galaxy. Galactic open clusters, with their well-defined single stellar populations, provide an ideal setting for exploring the helium spectral line feature. 
Red clump stars are specifically chosen for their constant surface helium abundance in a single stellar population, aiming to avoid potential intrinsic helium abundance dispersion induced by stellar evolution. 

The \He~profiles of our sample stars exhibit symmetric shapes and align with or around their rest wavelengths, indicating the absence of significant bulk motion in the upper chromosphere of these stars.
Blending lines of \ion{Ca}{i}, \ion{Fe}{i}, \ion{Si}{i}, \ion{Fe}{i}, and \ion{Ti}{i} in the \He~region are removed before measuring the helium line strength. Their abundances are determined from simultaneously observed optical high-resolution spectra, The oscillator strengths ($\loggf$) of these lines are calibrated ``astronomically''  during this process.

The final \He~line strengths of these red clump stars fall into two distinct categories. Three stars exhibit strong absorption, including a discernible secondary component related to the weaker line (at \SI{10829.09}{\angstrom}) of the \He~triplet, while the remaining stars display weaker absorption.
We identify a correlation between the main component of the \He~line strength and the \ion{Ca}{ii} $\logRpHK$ values. This result highlights the importance of accounting for stellar chromosphere structure when using the \He~line as an abundance indicator.

The method developed in this study sheds light on the possibility of using the \He~spectral line to probe stellar helium content, whenever simultaneous observations of other chromospheric diagnostics are available.  
The results can be further compared to helium measurements from future asteroseismology missions such as HAYDN \citep{Miglio2021}. 
We plan to extend this approach to additional open clusters and field stars spanning a broader range of Galactic regions and metallicities.
This will allow us to gain a more comprehensive understanding of helium abundance variations across different stellar populations and their implications for stellar and Galactic evolution.

\begin{acknowledgements}
This study is supported by JSPS KAKENHI Grant-in-Aid 21J11301.
X.F. thanks the support of the National Natural Science Foundation of China (NSFC) No. 12203100 and the China Manned Space Project with NO. CMS-CSST-2021-A08.
A.B. acknowledges funding from Mini-Grant INAF 2022 (High Resolution Observations of Open Clusters).
A.F. acknowledges funding from the Large-Grant INAF YODA (YSOs Outflow, Disks and Accretion).
The data of this study is based on observations made with the Italian Telescopio Nazionale Galileo (TNG) operated on the island of La Palma by the Fundación Galileo Galilei of the INAF (Istituto Nazionale di Astrofisica) at the Spanish Observatorio del Roque de los Muchachos of the Instituto de Astrofisica de Canarias. 
This research used the facilities of the Italian centre for Astronomical Archive (IA2) operated by INAF at the Astronomical Observatory of Trieste.
This work has made use of the VALD database, operated at Uppsala University, the Institute of Astronomy RAS in Moscow, and the University of Vienna.
\end{acknowledgements}

%
  \bibliographystyle{aa} 
  \bibliography{refs} 
%

\end{document}